\documentclass[]{svmult}
\usepackage{graphicx}
\usepackage{url}
\usepackage{amsmath,amssymb,amsfonts}
\usepackage{algorithmic}
\usepackage{algorithm}
\usepackage{array}
\usepackage[caption=false,font=normalsize,labelfont=sf,textfont=sf]{subfig}
\usepackage{booktabs}
\usepackage{xcolor}
\usepackage{multirow}
\usepackage{listings}
\usepackage{float}
\usepackage{newtxtext}
\usepackage{newtxmath}
\usepackage{makeidx}
\usepackage{multicol}
\usepackage{footmisc}

\usepackage{tikz}
\usetikzlibrary{arrows.meta, positioning, calc, shapes.geometric}

\definecolor{codegreen}{rgb}{0,0.6,0}
\definecolor{codegray}{rgb}{0.5,0.5,0.5}
\definecolor{codepurple}{rgb}{0.58,0,0.82}
\definecolor{backcolour}{rgb}{0.95,0.95,0.92}

\lstdefinestyle{mystyle}{
    backgroundcolor=\color{backcolour},   
    commentstyle=\color{codegreen},
    keywordstyle=\color{magenta},
    numberstyle=\tiny\color{codegray},
    stringstyle=\color{codepurple},
    basicstyle=\ttfamily\footnotesize,
    breakatwhitespace=false,         
    breaklines=true,                 
    captionpos=b,                    
    keepspaces=true,                 
    numbers=left,                    
    numbersep=5pt,                  
    showspaces=false,                
    showstringspaces=false,
    showtabs=false,                  
    tabsize=2
}

\begin{document}

\title*{Heuristic-Based Merging of HPC Traces to Extend  Hardware Counter Coverage}

\author{
Júlia Orteu~Aubach\orcidID{0009-0005-9121-6572}\and
Fabio Banchelli \orcidID{0000-0001-9809-0857}\and
Marc Clascà~Ramírez \orcidID{0000-0002-5963-6658}\and
Marta Garcia-Gasulla \orcidID{0000-0003-3682-9905}
}

\institute{
Julia Orteu~Aubach\at Barcelona Supercomputing Center (BSC), Barcelona, Spain \email{julia.orteu@bsc.es}\and
Fabio Banchelli \at Barcelona Supercomputing Center (BSC), Barcelona, Spain \email{fabio.banchelli@bsc.es} \and
Marc Clasca~Ramirez\at Barcelona Supercomputing Center (BSC), Barcelona, Spain \email{marc.clasca@bsc.es} \and
Marta Garcia-Gasulla\at Barcelona Supercomputing Center (BSC), Barcelona, Spain \email{marta.garcia@bsc.es}
}
\maketitle
\abstract{This work extends a framework for predicting the performance of High-Performance Computing (HPC) workloads using Machine Learning (ML).
A common limitation in performance modeling is the restricted number of hardware counters that can be collected simultaneously. 
To address this, we propose a heuristic-based methodology to merge execution traces from multiple runs, each instrumented with a different set of hardware counters. Our approach matches computation bursts across executions by analyzing MPI structure, timing, and communication patterns. 
This process enables the construction of a unified dataset that includes a wider set of hardware features without relying on multiplexing. 
The output is a new synthetic  trace  with all merged counters, which can be used both for HPC performance prediction and for conventional performance analysis. 
The methodology has been validated on MareNostrum5 machine with a range of kernels and real applications. 
Results show that the merged counters maintain acceptable accuracy depending on the application, and can be directly used to train ML models on a richer feature space without prior counter selection.
}


\section{Introduction}
\label{sec:introduction}

High-Performance Computing (HPC) applications present significant challenges for performance prediction due to the wide variety of factors that affect their behavior. These include the underlying hardware architecture, compiler optimizations, parallel programming paradigms (e.g., MPI, OpenMP), problem size, and the specific parallelization strategy. Moreover, HPC workloads span a broad range of domains, such as computational fluid dynamics (CFD), linear algebra solvers, seismic simulations, or molecular dynamics, each with distinct computational patterns and communication behaviors. These differences lead to highly diverse execution profiles, composed of computation and communication bursts with varying granularity, frequency, and resource demands. Such heterogeneity complicates the development of  generalizable performance models.

In previous work, we proposed a machine learning (ML) based framework to predict the performance of unseen HPC workloads \cite{Orteu_PPAM,tfg}. The approach relied on  execution traces with data read from hardware counters via the Performance Application Programming Interface (PAPI) library \cite{papi_website}  treating them as input features for ML models. However, an important limitation was the restricted number of counters that could be collected simultaneously, due to architectural constraints and counter incompatibilities. 

Traditionally, two main strategies exist to address this limitation: (i) selecting a reduced, fixed set of counters before execution, as done in our previous work, or (ii) employing hardware multiplexing to alternate counter sets at runtime. The first strategy limits feature diversity, while the second introduces  intermittent sampling.

In this work, we address this limitation by introducing a methodology to reconstruct a synthetic enriched trace that contains all desired counters. 
We propose a heuristic-based approach to merge multiple complete executions of the same application, each instrumented with different counter sets, by matching computation bursts across runs. This merging process enables the construction of a unified trace with extended counter coverage per burst, expanding the feature space available for training ML models. 

The remainder of this article is organized as follows: Section \ref{sec:related_work} discusses related work, Section \ref{sec:methodology} presents our methodology, Section \ref{sec:experimental_setup} details the experimental setup, Section \ref{sec:results} provides results and analysis, and Section \ref{sec:conclusion} offers conclusions and future work.

\section{Related Work}
\label{sec:related_work}

\begin{figure}[h!tbp]
\centering
\includegraphics[width=1\columnwidth]{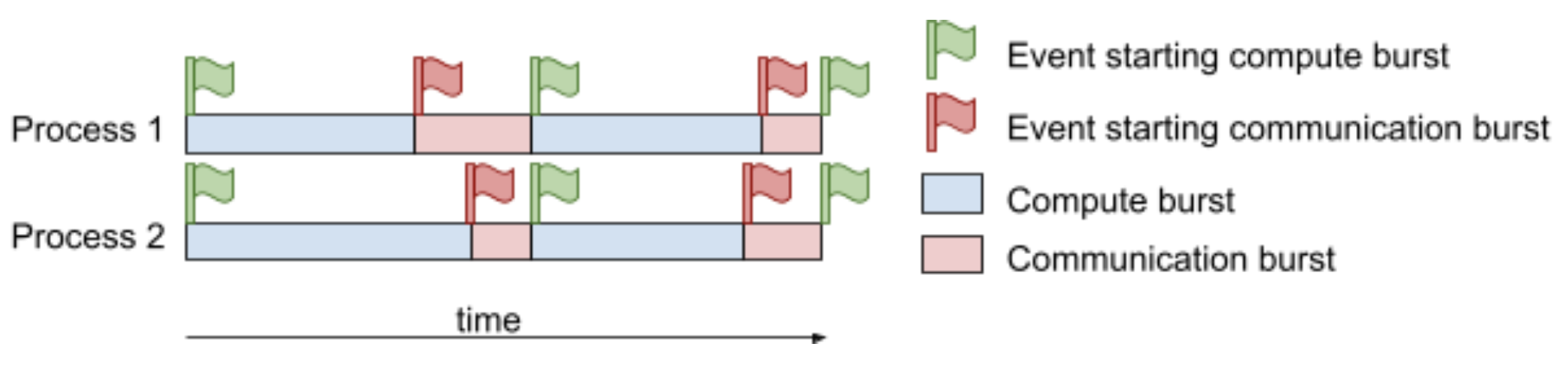}
\caption{\textit{Extrae} and \textit{Paraver} generalization of burst.}
\label{fig:trace}  
\end{figure}

Tracing-based tools, such as \textit{Extrae} and \textit{Paraver}, are widely used to capture and inspect the fine-grained behavior of parallel applications. \textit{Extrae}  is a software package that automatically instruments HPC applications by intercepting function calls from the most common parallel programming models (e.g., MPI, OpenMP, CUDA), collects hardware-counter values through PAPI, and produces a time-stamped event stream per thread or process generating post-mortem execution traces \cite{extrae}. A burst is the closed interval between any two successive events in that stream; in practice, we distinguish compute bursts (the region between two MPI calls) from communication bursts (the MPI call itself). Figure \ref{fig:trace} is a drawing of this event-to-burst generalization.

\textit{Paraver} renders these traces as timelines and statistics for burst-level analysis that guide manual analysis \cite{paraver}. González \textit{et al.} first automated that inspection by applying the DBSCAN algorithm to the set of hardware counters collected for every compute burst,  clustering recurring computation phases without user supervision \cite{clustering}. Their technique assumes, however, that all counters of interest are present in a single execution; an assumption that fails on many architectures, where only a handful of counters can be recorded concurrently and numerous combinations are disallowed.

Our previous framework used decision tree ensemble models trained on \textit{Extrae/Paraver} traces to predict the performance of unseen HPC workloads \cite{doctoral_poster_julia_2024, Orteu_PPAM, tfg}. In practice the predictive power of those models is capped by the \textit{“n-counter ceiling’}’: because only a small, fixed set of counters can be recorded due to imposed by architectural incompatibilities, limiting the expressiveness of the feature space. 

The limitations of hardware performance counters (HWC) have been  documented in the literature, for instance Das \textit{et al.} \cite{das} conducted a study of HWC  usage across nearly 100 papers, finding that modern processors typically provide only 4-8 concurrent counters despite offering hundreds of monitorable events. Their analysis reveals  issues with non-determinism and overcounting, showing that identical executions can produce counter variations of 1-5\%.  Azimi \textit{et al.} \cite{azimi} address the limited counter availability through statistical multiplexing, alternating counter groups at fine time intervals to provide more logical counters. While their approach achieves reasonable accuracy, it inherently compromises temporal precision by sampling counter groups intermittently rather than continuously.

In the literature several strategies have been explored to enlarge that space of HWC, none fully satisfactory for fine-grained, burst-oriented analysis:

\begin{itemize}
    \item Hardware multiplexing alternates counter groups at runtime. While this increases coverage, it samples each group intermittently, introduces timing noise due to the overhead of switching between counter sets, and still fails to provide complete streams for every burst.
    \item Multi-run profile fusion (e.g., HPCToolkit’s post-mortem merge \cite{zhou21}) aggregates metrics from separate runs into a single calling-context tree, yet loses the temporal alignment required by \textit{Paraver}.
    \item Cross-run event alignment algorithms such as the one proposed by Pandey \textit{et al.}  meta-graph matching \cite{pandey23}  match logical events across traces but ignore hardware counters; they do not attempt to combine separate counter streams. 
\end{itemize}

The approach proposed in this study aims to overcome that limitation and expand our feature space by recording several full executions of the same application, each instrumented with a complementary subset of counters. An MPI-aware heuristic aligns corresponding compute bursts across runs and produces a synthetic trace that contains every counter for every burst while preserving the original temporal structure. \\

To our knowledge, this is the first burst-level fusion method that extends hardware counter coverage without sacrificing the fidelity required for traditional Paraver analysis.

\section{Methodology}
\label{sec:methodology}

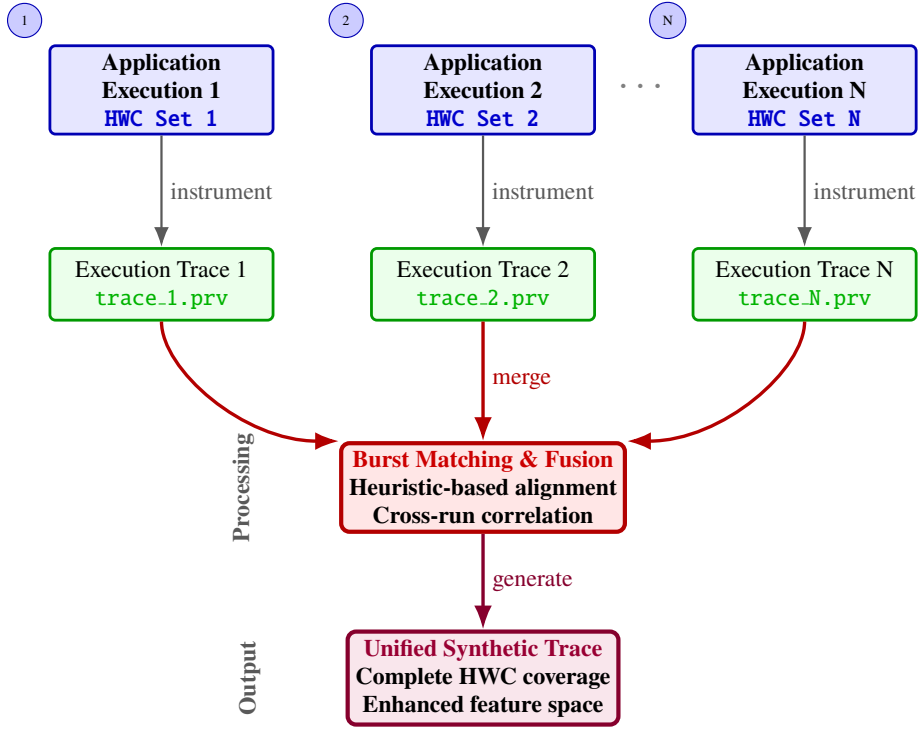
\begin{figure}[htbp]
\centering
\resizebox{\columnwidth}{!}{%
\begin{tikzpicture}[
    node distance=1.2cm and 1.2cm,
    execbox/.style={rectangle, draw=blue!70!black, fill=blue!10, rounded corners=2pt, 
                    minimum width=2.8cm, minimum height=0.9cm, align=center, 
                    line width=1pt, font=\footnotesize\bfseries},
    tracebox/.style={rectangle, draw=green!60!black, fill=green!8, rounded corners=2pt, 
                     minimum width=2.8cm, minimum height=0.9cm, align=center, 
                     line width=1pt, font=\footnotesize},
    fusionbox/.style={rectangle, draw=red!70!black, fill=red!10, rounded corners=3pt, 
                      minimum width=3.2cm, minimum height=1.1cm, align=center, 
                      line width=1.5pt, font=\footnotesize\bfseries},
    resultbox/.style={rectangle, draw=purple!70!black, fill=purple!10, rounded corners=3pt, 
                      minimum width=3.2cm, minimum height=1.1cm, align=center, 
                      line width=1.5pt, font=\footnotesize\bfseries},
    arrow/.style={-{Latex[length=2.5mm]}, thick, color=gray!70!black},
    fusionarrow/.style={-{Latex[length=3mm]}, very thick, color=red!70!black},
    every node/.style={font=\footnotesize}
]

\node[execbox] (exec1) {Application\\Execution 1\\{\color{blue!80!black}\texttt{HWC Set 1}}};
\node[execbox, right=of exec1] (exec2) {Application\\Execution 2\\{\color{blue!80!black}\texttt{HWC Set 2}}};
\node[execbox, right=of exec2] (exec3) {Application\\Execution N\\{\color{blue!80!black}\texttt{HWC Set N}}};

\node[font=\Large\bfseries, color=gray] at ($(exec2.east)!0.5!(exec3.west)$) {$\cdots$};

\node[tracebox, below=1.4cm of exec1] (trace1) {Execution Trace 1\\{\color{green!70!black}\texttt{trace\_1.prv}}};
\node[tracebox, below=1.4cm of exec2] (trace2) {Execution Trace 2 \\{\color{green!70!black}\texttt{trace\_2.prv}}};
\node[tracebox, below=1.4cm of exec3] (trace3) {Execution Trace N\\{\color{green!70!black}\texttt{trace\_N.prv}}};

\draw[arrow] (exec1) -- (trace1) node[midway, right, font=\footnotesize] {instrument};
\draw[arrow] (exec2) -- (trace2) node[midway, right, font=\footnotesize] {instrument};
\draw[arrow] (exec3) -- (trace3) node[midway, right, font=\footnotesize] {instrument};

\node[fusionbox, below=1.5cm of trace2] (fusion) {{\color{red!80!black}Burst Matching \& Fusion}\\{\footnotesize Heuristic-based alignment}\\{\footnotesize Cross-run correlation}};

\draw[fusionarrow] (trace1.south) .. controls +(down:0.6cm) and +(left:1.2cm) .. (fusion.north west);
\draw[fusionarrow] (trace2.south) -- (fusion.north) node[midway, right, font=\footnotesize] {merge};
\draw[fusionarrow] (trace3.south) .. controls +(down:0.6cm) and +(right:1.2cm) .. (fusion.north east);

\node[resultbox, below=of fusion] (result) {{\color{purple!80!black}Unified Synthetic Trace}\\{\footnotesize Complete HWC coverage}\\{\footnotesize Enhanced feature space}};

\draw[fusionarrow, color=purple!70!black] (fusion) -- (result) node[midway, right, font=\footnotesize] {generate};


\node[rotate=90, anchor=center, font=\footnotesize\bfseries, color=gray!70!black] 
      at ([xshift=-1.2cm]fusion.west) {Processing};
\node[rotate=90, anchor=center, font=\footnotesize\bfseries, color=gray!70!black] 
      at ([xshift=-1.2cm]result.west) {Output};

\node[circle, fill=blue!20, draw=blue!70!black, minimum size=4mm, font=\tiny] 
      at ([xshift=-0.3cm, yshift=0.3cm]exec1.north west) {1};
\node[circle, fill=blue!20, draw=blue!70!black, minimum size=4mm, font=\tiny] 
      at ([xshift=-0.3cm, yshift=0.3cm]exec2.north west) {2};
\node[circle, fill=blue!20, draw=blue!70!black, minimum size=4mm, font=\tiny] 
      at ([xshift=-0.3cm, yshift=0.3cm]exec3.north west) {N};

\end{tikzpicture}%
}
\caption{Methodology overview for trace fusion: Multiple application executions with disjoint hardware counter sets are instrumented to generate individual traces, which are subsequently merged through a burst matching heuristic  to produce a unified synthetic trace with extended counter coverage.}
\label{fig:trace_fusion_methodology}
\end{figure}

The methodology consists of three main stages: (A) data collection across $N$ executions, each one with a unique counter set, (B) burst matching using MPI-aware heuristics, and (C) trace fusion to create synthetic enriched traces. 
Figure~\ref{fig:trace_fusion_methodology} provides an overview of this process. 
Currently, the methodology has been applied \emph{to MPI-only applications executed on CPUs}. Future work includes extending the methodology to hybrid parallelization models and heterogeneous hardware environments.

\vspace{-0.4cm}
\subsection{Data Collection}
\label{subsec:data_collection}

Multiple execution traces are collected from the same HPC application binary, where each run is instrumented through \textit{Extrae} with a different set of PAPI counters\footnote{To ensure counter compatibility, \textit{Extrae} provides the \texttt{papi\_best\_set} tool. This binary allows users to specify all desired performance counters and automatically checks their availability on the target machine. It then organizes compatible counters into optimal sets, indicating which counters can be measured simultaneously and which would require separate sets with multiplexing. For this study, only compatible counter combinations within each set are used to ensure simultaneous measurement without using multiplexing}.

The trace processing pipeline consists of several stages that transform raw \textit{Paraver} traces into structured burst-level datasets. The extraction process employs parallel processing to handle multiple configuration files simultaneously, reducing processing time. The core extraction function processes each trace file through the following sequence:

\begin{enumerate}

    \item \textbf{Raw Data Gathering}: A custom parser developed in prior work \cite{Orteu_PPAM} is used to convert \texttt{.prv} traces into a burst-level tabular format (CSV), where each row corresponds to a burst and each column represents a feature. This includes all counters recorded per burst, both communication and computation. The basic data collected includes \texttt{TaskId}, \texttt{Begin\_Time}, \texttt{End\_Time}, \texttt{Duration}, and the specific hardware counters from the configured set.

    \item \textbf{Contextual Feature Extraction}: \textit{Paraver} allows applying configuration files with custom rules to post-process traces. In our work, we use this mechanism to extract additional contextual metrics from the trace data:
        \begin{itemize}
            \item \textbf{MPI Context}: Extract MPI call types, communication partners, and message sizes for each communication burst to characterize the communication behavior surrounding each computation burst. This configuration-based approach is flexible and can be extended to extract other types of contextual information from the same traces.
        \item \textbf{Concurrency Metrics}: Calculate average process concurrency levels to quantify how many processes were simultaneously active on each compute node during the execution of each burst, capturing the degree of resource-level parallelism.

        \end{itemize}
    
        \item \textbf{Derived Metrics}: Calculate features such as burst frequency, instructions per cycle (IPC), and the relative position of each burst within its process trace (i.e., the percentage of bursts completed at that point in the execution). These are computed from the extracted raw counters and contextual information.
    
    \item \textbf{Data Consolidation}: Merge all extracted features into a unified dataset with proper temporal and process alignment.
\end{enumerate}

The resulting dataset contains approximately 20-25 features per counter set, combining temporal information, hardware counters, concurrency metrics, and derived features. In particular, all MPI structural information is augmented with \texttt{\_before} and \texttt{\_after} variants to characterize the communication context surrounding each computation burst. This approach captures the nature of communication events that precede and follow computational phases, providing 
contextual signatures for each burst.

\subsection{Observed Communication Patterns Across Applications}
\label{subsec:patterns_observed}

Throughout the evaluation of our methodology on a diverse set of HPC applications, we observed three distinct behaviors in terms of communication structure across MPI processes:

\begin{itemize}
    \item \textbf{Processes with identical communication behavior}: All MPI ranks exhibit the same number of computation bursts and follow the same MPI communication sequences. These cases allow for direct, position-based burst matching.
    
    \item \textbf{Processes with structural variations}: MPI ranks present different communication patterns or burst counts, but maintain recurring structures (e.g., specific MPI sequences that repeat with consistent frequency), enabling pattern-based alignment.
    
    \item \textbf{Highly irregular traces}: Some applications exhibit complex and heterogeneous communication behavior, with significant structural irregularities across processes and executions. These cases require a more sophisticated matching strategy that considers structural constraints and contextual features.
\end{itemize}

Identifying these scenarios guided the design of our two-stage matching algorithm, which progressively increases the complexity of matching heuristics based on the level of structural similarity detected.


\subsection{Heuristic-Based Trace Merging}
\label{subsec:trace_merging}

\begin{figure}[htbp]
\centering
\resizebox{\columnwidth}{!}{%
\begin{tikzpicture}[
    node distance=1.2cm and 0.8cm,
    inputbox/.style={rectangle, draw=blue!70!black, fill=blue!8, rounded corners=2pt, 
                     minimum width=2.2cm, minimum height=0.7cm, align=center, 
                     line width=1pt, font=\footnotesize\bfseries},
    stage1box/.style={rectangle, draw=teal!70!black, fill=teal!10, rounded corners=2pt, 
                      minimum width=2.6cm, minimum height=0.9cm, align=center, 
                      line width=1.5pt, font=\footnotesize\bfseries},
    stage1detail/.style={rectangle, draw=teal!60!black, fill=teal!5, rounded corners=2pt, 
                         minimum width=2.8cm, minimum height=0.8cm, align=center, 
                         line width=1pt, font=\footnotesize},
    stage2box/.style={rectangle, draw=orange!70!black, fill=orange!12, rounded corners=2pt, 
                      minimum width=2.6cm, minimum height=0.9cm, align=center, 
                      line width=1.5pt, font=\footnotesize\bfseries},
    stage2detail/.style={rectangle, draw=orange!60!black, fill=orange!8, rounded corners=2pt, 
                         minimum width=2.8cm, minimum height=0.8cm, align=center, 
                         line width=1pt, font=\footnotesize},
    decisionbox/.style={diamond, draw=purple!70!black, fill=purple!10, 
                        minimum width=1.8cm, minimum height=0.7cm, align=center, 
                        line width=1pt, font=\footnotesize\bfseries},
    outputbox/.style={rectangle, draw=green!70!black, fill=green!10, rounded corners=2pt, 
                      minimum width=2.4cm, minimum height=0.7cm, align=center, 
                      line width=1pt, font=\footnotesize\bfseries},
    arrow/.style={-{Latex[length=2mm]}, thick, color=gray!70!black},
    stage1arrow/.style={-{Latex[length=2mm]}, thick, color=teal!70!black},
    stage2arrow/.style={-{Latex[length=2mm]}, thick, color=orange!70!black},
    yesarrow/.style={-{Latex[length=2mm]}, thick, color=red!70!black},
    noarrow/.style={-{Latex[length=2mm]}, thick, color=green!70!black},
    every node/.style={font=\footnotesize}
]

\node[inputbox] (input1) {Trace 1\\{\color{blue!80!black}\footnotesize Unmatched bursts}};
\node[inputbox, right=1.2cm of input1] (input2) {Trace 2\\{\color{blue!80!black}\footnotesize Unmatched bursts}};
\node[inputbox, right=1.2cm of input2] (inputN) {Trace N\\{\color{blue!80!black}\footnotesize Unmatched bursts}};

\node[stage1box, below=1.8cm of input2] (stage1) {{\color{teal!90!black}STAGE 1}\\{\footnotesize Primary Matching}\\{\footnotesize Parallel by MPI rank}};

\draw[stage1arrow] (input1.south) .. controls +(down:0.7cm) and +(left:0.8cm) .. (stage1.north west);
\draw[stage1arrow] (input2.south) -- (stage1.north);
\draw[stage1arrow] (inputN.south) .. controls +(down:0.7cm) and +(right:0.8cm) .. (stage1.north east);

\node[stage1detail, right=1.8cm of stage1, minimum width=3.0cm] (direct) {
    {\color{teal!90!black}\textbf{Direct Matching}}\\
    {\footnotesize IF: Identical burst count}\\
    {\footnotesize AND: Identical MPI sequences}\\
    {\footnotesize → Sequential assignment}
};

\node[stage1detail, below=0.5cm of direct, minimum width=3.0cm] (pattern) {
    {\color{teal!90!black}\textbf{Pattern-Based Matching}}\\
    {\footnotesize ELSE: MPI frequency analysis}\\
    {\footnotesize Consistent patterns across runs}\\
    {\footnotesize → Match by temporal order}
};

\node[decisionbox, below=1.3cm of stage1] (decision) {{\color{purple!80!black}Unmatched}\\{\color{purple!80!black}bursts?}};
\draw[arrow] (stage1) -- (decision);

\node[stage2box, below=1.3cm of decision] (stage2) {{\color{orange!90!black}STAGE 2}\\{\footnotesize Remaining Bursts}\\{\footnotesize Structural Constraints}};
\draw[yesarrow] (decision.south) -- (stage2.north) node[midway, right, font=\small] {\textbf{Yes}};

\node[stage2detail, right=1cm of stage2, minimum width=3.0cm] (structural) {
    {\color{orange!90!black}\textbf{Structural Matching}}\\
    {\footnotesize 1. Collective Region Definition}\\
    {\footnotesize 2. MPI Structure-Based Grouping}\\
    {\footnotesize 3. Multi-Criteria Similarity Matching}\\
    {\footnotesize \textit{(weighted function)}}

};

\node[outputbox, below=1.1cm of stage2] (final) {{\color{green!80!black}Matched Bursts}\\{\footnotesize with burst\_id}};
\draw[stage2arrow] (stage2) -- (final);

\draw[noarrow] (decision.west) .. controls +(left:3cm) and +(left:2cm) .. (final.north west) node[midway, above left, font=\small] {\textbf{No}};

\draw[arrow, dashed, color=teal!60!black] (stage1.east) .. controls +(right:1.0cm) and +(left:0.4cm) .. (direct.west);
\draw[arrow, dashed, color=teal!60!black] (stage1.east) .. controls +(right:1.0cm) and +(left:0.4cm) .. (pattern.west);
\draw[arrow, dashed, color=orange!60!black] (stage2.east) -- (structural.west);

\node[circle, fill=teal!20, draw=teal!70!black, minimum size=5mm, font=\small\bfseries] 
      at ([xshift=-0.3cm, yshift=0.5cm]stage1.north west) {1};
\node[circle, fill=orange!20, draw=orange!70!black, minimum size=5mm, font=\small\bfseries] 
      at ([xshift=-0.3cm, yshift=0.3cm]stage2.north west) {2};

\end{tikzpicture}%
}
\caption{Two-stage burst matching algorithm: \textcolor{teal!80!black}{\textbf{Stage 1}} performs direct and pattern-based matching for consistent communication patterns, while \textcolor{orange!80!black}{\textbf{Stage 2}} applies structural constraint-based matching using collective operation boundaries and weighted similarity functions to generate a complete synthetic trace.}
\label{fig:burst_matching_algorithm}
\end{figure}
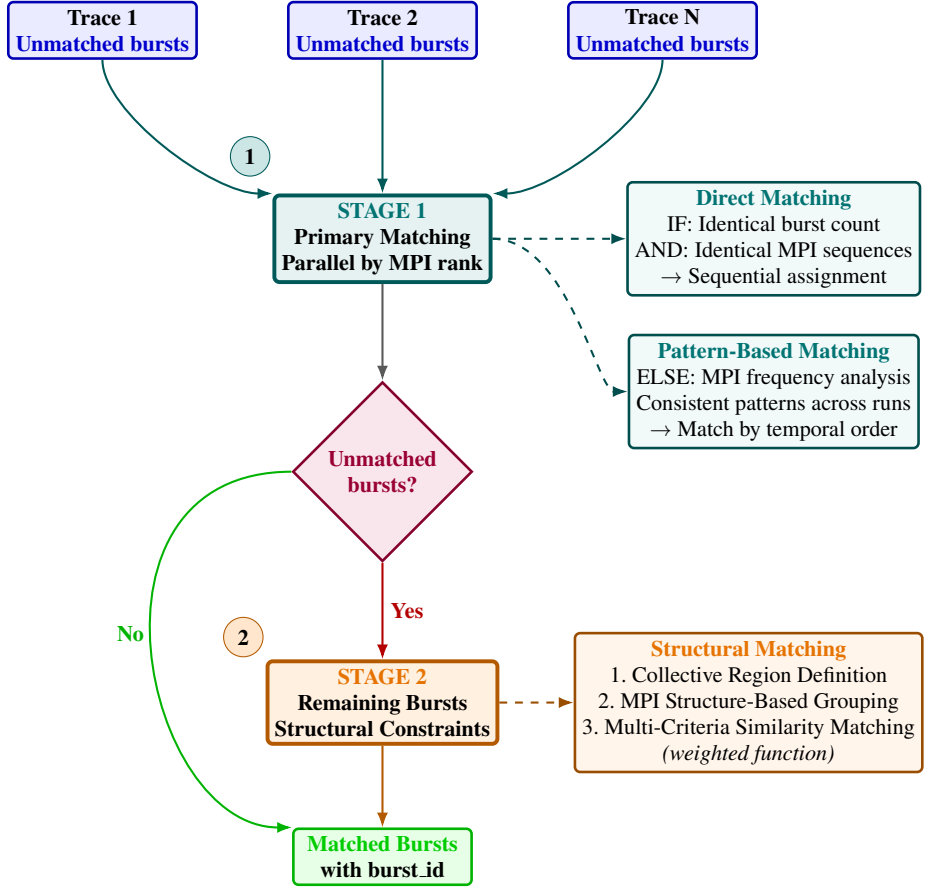

The burst matching process operates through a two-stage approach that leverages the nature of MPI communication patterns to identify corresponding computation bursts across different executions. The algorithm exclusively focuses on computation bursts (referred to as "\textit{useful bursts}" in alignment with our previous work for performance prediction), filtering out communication bursts while using their stored communication context to guide the matching process. The decision graph in figure \ref{fig:burst_matching_algorithm} depicts the two stages of the algorithm, from individual traces to a set of matched bursts. Since all applications are executed with identical process counts, the algorithm processes each MPI rank independently in parallel.

\subsubsection{Stage 1 - Primary Burst Matching}

The first stage employs a dual approach: \emph{Direct Matching} for processes with identical communication behavior (same burst count and MPI sequences across all executions), where bursts are matched by temporal position; and \emph{Pattern-Based Matching} for processes with structural variations, which identifies recurring MPI patterns with consistent frequency across runs (e.g., if \texttt{MPI\_ALLREDUCE} followed by \texttt{MPI\_BCAST} occurs exactly 15 times in all executions, these structurally equivalent occurrences are matched by temporal order). The algorithm for this stage is depicted in Algorithm \ref{app:alg} in the Appendix.

\subsubsection{Stage 2 - Remaining Burst Matching with Structural Constraints}

Bursts that remain unmatched after Stage 1 undergo  a three-phase approach that respects MPI collective communication boundaries while processing each MPI rank in parallel.

\begin{enumerate}
    \item \textbf{Collective Region Definition.} The algorithm leverages MPI collective operations (\texttt{MPI\_BARRIER}, \texttt{MPI\_BCAST}, \texttt{MPI\_ALLREDUCE}, \texttt{MPI\_GATHER}, \texttt{MPI\_SCATTER}, and variants) as natural synchronization points that divide execution into coherent regions. Each region is defined as the temporal interval $(t_{start}, t_{end}, region_{id})$ between consecutive collectives, ensuring bursts are matched only within structurally equivalent execution phases.
    \item \textbf{MPI Structure-Based Grouping.} Within each region, unmatched bursts are grouped by their communication context $(MPI_{before}, MPI_{after})$. Groups appearing in fewer than two executions are filtered out, and each group is identified by the composite key $(MPI_{structure}, region_{id})$.

    \item \textbf{Multi-Criteria Similarity Matching.} Bursts within each group are matched using a weighted similarity function combining three features:
        \begin{itemize}
        \item \textit{Temporal position}: Relative position within the collective region ($D_{temporal}$).
        \item \textit{Communication size}: Message sizes for MPI operations before and after the burst, computed as:  $$D_{size} = \frac{|size_{before}^{ref} - size_{before}^{candidate}|}{\max(size_{before}^{ref}, size_{before}^{candidate}, 1)}$$ 
        \item \textit{Communication partner}: Rank identifier for point-to-point communication ($D_{partner} = 0$ if partners match, $1$ otherwise).
        \end{itemize}
        The similarity score between the reference and candidate bursts is:  
        \begin{center}
                    $S = 0.6 \cdot D_{temporal} + 0.2 \cdot D_{size} + 0.2 \cdot D_{partner}$

        \end{center}
        
     Bursts are matched only if $S < 0.3$, ensuring matches represent genuinely similar computational contexts.
\end{enumerate}




\subsection{Trace Fusion}
\label{subsec:trace_fusion}

Once burst correspondences are established across executions, the final stage creates a unified synthetic trace that combines hardware counters from all runs while preserving temporal structure. The fusion process operates exclusively on successfully matched computation bursts, discarding unmatched bursts to ensure data integrity.

The algorithm selects the execution with the lowest unmatched burst rate as the base trace to maximize data completeness. For each additional execution, columns are merged using the following strategy:

\begin{itemize}
    \item \textbf{Identical columns}: Retain single version if values are equivalent across executions.
    \item \textbf{Divergent columns}: Add with execution-specific prefixes (e.g., \texttt{run2\_PAPI\_L1\_DCM}).
    \item \textbf{Unique columns}: Merge directly with appropriate prefixes to distinguish counter sets.
\end{itemize}

The resulting synthetic trace maintains the original temporal sequence (\texttt{TaskId}, \texttt{Begin\_Time,} \texttt{Duration}) from the base execution while providing access to the complete set of hardware counters that would be impossible to collect in a single execution due to architectural constraints. 

\section{Experimental Setup}
\label{sec:experimental_setup}

This section describes the experimental methodology used to evaluate the proposed trace merging approach, including the execution environment, benchmark applications, hardware counter configurations and validation strategy.

 \vspace{-0.5cm}
\subsection{Execution Environment and Parameters}
The experiments are conducted on MareNostrum5, the pre-exascale system located at BSC \cite{MN5}. For this project, only the general-purpose partition is used. It is based on Intel Sapphire Rapids CPUs.

Throughout the study, the following configuration was used:

\begin{itemize}
    \item \textbf{CPU Frequency}: Fixed at 2 GHz to minimize timing variations.
    \item \textbf{Node Allocation}: Exclusive node access to reduce system noise.
    \item \textbf{MPI Processes}: Varies by application (typically 100-112 processes).
    \item \textbf{Repetitions}: 2-10 runs per application and counter set combination.
\end{itemize}

\vspace{-0.5cm}

\subsection{Hardware Counter Sets}

The experimental setup employs three complementary counter sets designed to capture distinct aspects of application behavior while respecting MareNostrum5's architectural compatibility constraints. The counter selection is guided by relevance for future ML-based performance prediction models, with each set targeting specific computational characteristics. The \emph{INS\_MIX} set focuses on memory hierarchy behavior and instruction composition, capturing cache miss patterns and instruction mix. The \emph{OPS\_SET} emphasizes floating-point operations and vectorization efficiency and the \emph{OPS\_CYC} set combines cycle and operation counts to enable direct comparison and validation against the other two sets. Table~\ref{tab:counter_sets} details the specific PAPI counters included in each set.

\begin{table}[h!]
\centering
\caption{Hardware Counter Sets Configuration}
\label{tab:counter_sets}
\begin{tabular}{|l|l|l|}
\hline
\textbf{Set Name} & \textbf{Focus} & \textbf{PAPI Counters} \\
\hline
\multirow{3}{*}{INS\_MIX} & Instruction Mix & PAPI\_TOT\_INS, PAPI\_TOT\_CYC, \\
& and Cache Hierarchy & PAPI\_LD\_INS, PAPI\_SR\_INS, \\
& & PAPI\_BR\_INS, PAPI\_L3\_TCM, \\
& & PAPI\_L1\_DCM, PAPI\_L2\_DCM \\
\hline
\multirow{3}{*}{OPS\_SET} & Floating-Point & PAPI\_TOT\_INS, PAPI\_VEC\_INS, \\
& Operations and & PAPI\_FP\_INS, PAPI\_FP\_OPS, \\
& Vectorization & PAPI\_DP\_OPS, PAPI\_SP\_OPS, \\
& & PAPI\_VEC\_SP, PAPI\_VEC\_DP \\
\hline
\multirow{2}{*}{OPS\_CYC} & Computational & PAPI\_TOT\_INS, PAPI\_TOT\_CYC, \\
& Performance & PAPI\_VEC\_DP, PAPI\_VEC\_SP, \\
& and Cycles & PAPI\_DP\_OPS \\
\hline
\end{tabular}
\end{table}

\subsection{Benchmark Applications}
\label{sec:benchamrks}
Our evaluation uses five representative HPC applications that span different computational domains and exhibit distinct performance characteristics. Table~\ref{tab:applications} summarizes the applications used in this study.

\begin{table}[h!]
\centering
\caption{HPC Applications Used for Evaluation}
\label{tab:applications}
\begin{tabular}{|c|p{2cm}|p{4cm}|}
\hline
\textbf{Application} & \textbf{Domain} & \textbf{Description} \\
\hline
SOD2D & Computational Fluid Dynamics & Spectral high-Order code for solving partial differential equations, primarily used for fluid dynamics and wave propagation problems \cite{SOD2D}.\\
\hline
SeisSol & Seismic Simulation & High-performance seismic wave simulation software for earthquake modeling and ground motion prediction \cite{SeisSol}.\\
\hline
Stream & Memory Benchmark & Memory bandwidth benchmark measuring performance of four vector operations: Copy, Scale, Add, and Triad \cite{stream}. \\
\hline
Alya Solver & Computational Mechanics & Mini-app from HPC mechanics application featuring very fine-grained parallelism for computational mechanics problems \cite{alya}. \\
\hline
Lulesh & Hydrodynamics & Benchmark tool for shock wave simulations in fluid dynamics, focusing on efficient energy calculations for nuclear fission explosions \cite{LULESH2:changes}. \\
\hline
\end{tabular}
\end{table}

\subsection{Evaluation Methodology}
\label{subsec:evaluation_methodology}
Since multiple executions are collected for each application and counter set combination, the same heuristic matching algorithm (Section~\ref{subsec:trace_merging}) can be applied to traces with identical counter sets to assess matching precision. This approach provides a controlled method to quantify matching errors by comparing identical hardware counters across matched bursts, establishing the baseline accuracy of the burst correspondence identification.

\subsubsection{Quality Assessment Framework}

For each application and counter set, N executions are processed through the matching algorithm. Quality is quantified using three metrics:

\begin{itemize}

\item \textbf{Pearson Correlation:} Correlation coefficient between base trace and each matched trace. Values $>$ 0.95 indicate high matching accuracy.

 \item \textbf{Relative Difference:} This metric quantifies the deviation between the base trace and the average of all other matched traces. Let $b_i$ be the base value for burst $i$, and $\mu_i=\frac{1}{N-1}\sum_{j\neq base} t_{j,i}$ the mean across the $N-1$ matched traces (excluding the base). We compute it only on rows where $b_i>0$:
 \[
 \mathrm{RelDiff} \;=\; \frac{1}{|\mathcal{I}|}\sum_{i\in\mathcal{I}} \frac{|\,b_i-\mu_i\,|}{b_i},
 \qquad \mathcal{I}=\{\,i:\; b_i>0\,\}.
 \]
     \item \textbf{Mean Absolute Error (MAE):} Absolute deviation (in original units) of the base trace from the average of the other matched traces, including zeros:
 \[
 \mathrm{MAE} \;=\; \frac{1}{M}\sum_{i=1}^{M} \bigl|\,b_i-\mu_i\,\bigr|,
 \]
 where $M$ is the number of the matched bursts.

\end{itemize}

To ensure a fair evaluation while removing only extreme outliers, we apply a one-sided fence based on the central $90\%$ range to the per-burst scores  \emph{before} computing, for each metric, the mean across all bursts values. Let $q_{0.05}$ and $q_{0.95}$ denote the 5th and 95th percentiles of the per-burst scores. We set
\[
U = q_{0.95} + 1.5\,\bigl(q_{0.95}-q_{0.05}\bigr),
\]
and keep values $\le U$. This conservative choice removes only the most extreme mismatches.

\subsubsection{Validation Criteria}

The assessment analyzes all common features (HWC, temporal data, communication information and relative position) across matched bursts, computing correlation matrices and relative difference distributions. Additional metrics include the percentage of bursts with relative differences below 30\% as a global quality indicator.

Visual analysis through scatter plots, correlation matrices, and distribution comparisons provides additional validation evidence for the subsequent results section.

\section{Results and Analysis}

\label{sec:results}

The burst matching methodology was evaluated across the different HPC applications detailed in Section~\ref{sec:benchamrks}, each presenting distinct computational patterns and MPI communication structures. Results demonstrate varying matching success rates depending on application characteristics.

\subsection{Direct Matching Applications}

\begin{figure}[h!tbp]
\centering
\includegraphics[width=1\columnwidth]{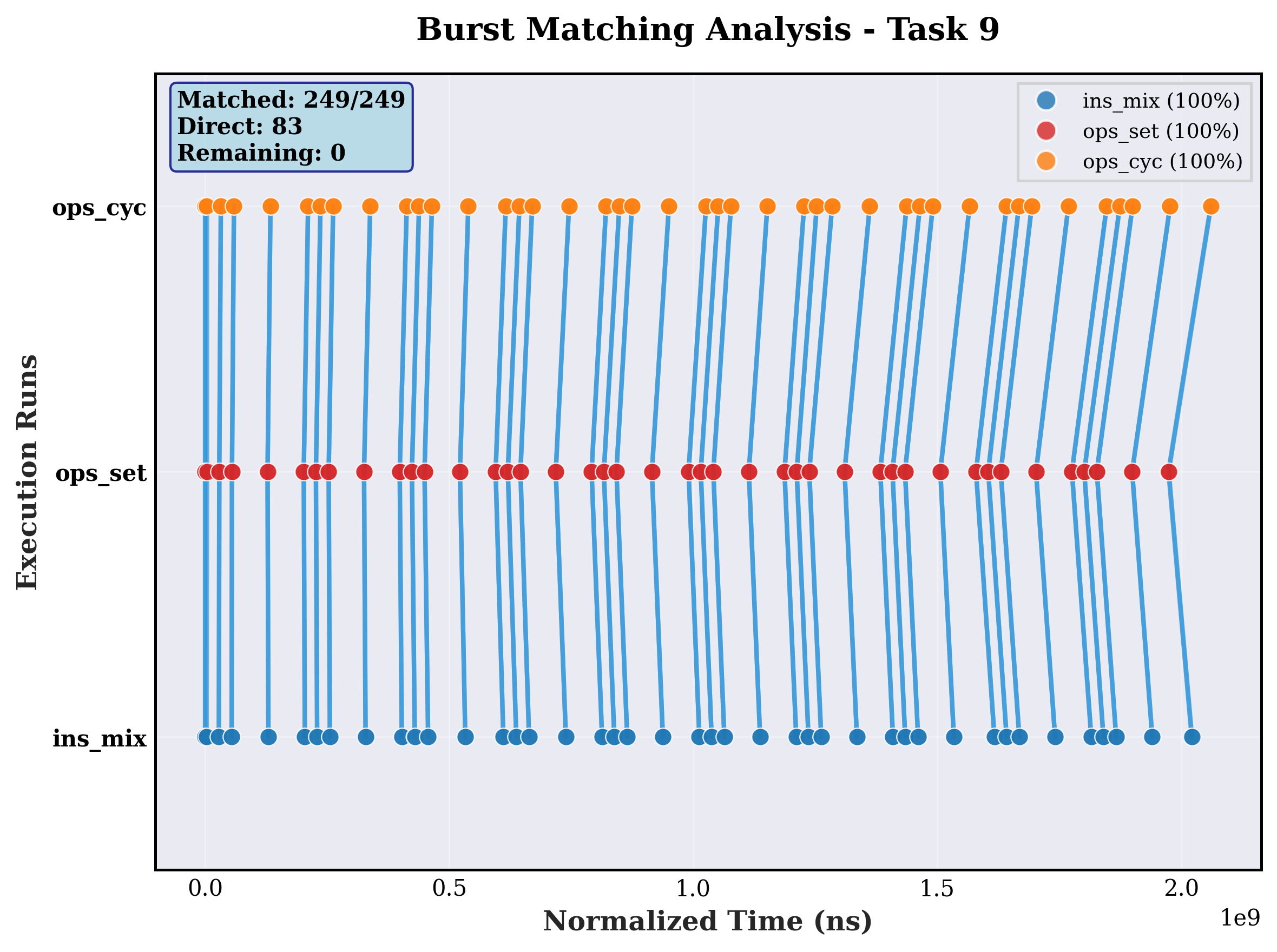}
\caption{Temporal burst alignment in \textbf{Stream} application for TaskId 9. Blue connecting lines show perfect correspondence between bursts across three counter configurations, with each colored circle representing bursts from different executions. The regular temporal pattern and absence of unmatched bursts demonstrates the deterministic nature enabling direct matching success.}
\label{fig:stream_matching}
\end{figure}

Three applications (Stream, Alya, Lulesh) achieved perfect 100\% burst matching through the direct matching algorithm (see Table \ref{tab:direct_matching}). Stream exhibited identical burst counts and MPI context per TaskId across executions, while Alya and Lulesh maintained identical MPI structure throughout the entire trace in each execution, enabling straightforward temporal alignment. Figure~\ref{fig:stream_matching} shows an example of the direct  alignment that confirms the reliability of the direct matching approach for deterministic applications.

\begin{table}[h!]
\centering
\caption{Direct Matching Performance Summary}
\label{tab:direct_matching}
\begin{tabular}{|l|r|c|c|r|}
\hline
\textbf{Application} & \textbf{Bursts per Exec Set} & \textbf{Match Rate} & \textbf{Method} & \textbf{TaskIds} \\
\hline
Stream & 9,296 & 100\% & Direct & 112 \\
Alya & 16,428 & 100\% & Direct & 48 \\
Lulesh & 43,077 & 100\% & Direct & 27 \\
\hline
\end{tabular}
\end{table}
\vspace{-0.2cm}

\subsection{Pattern-Based Matching Applications}
Applications with structural variations between executions required advanced pattern-based matching algorithms. SOD2D and SeisSol presented different levels of matching complexity due to execution non-determinism and varying communication patterns across runs.

\subsubsection{SOD2D: High-Success Two-Stage Matching}
The application exhibited minor structural variations across executions, with burst counts differing by only 0.02\% between counter sets.

\begin{figure}[h!tbp]
\centering
\includegraphics[width=1\columnwidth]{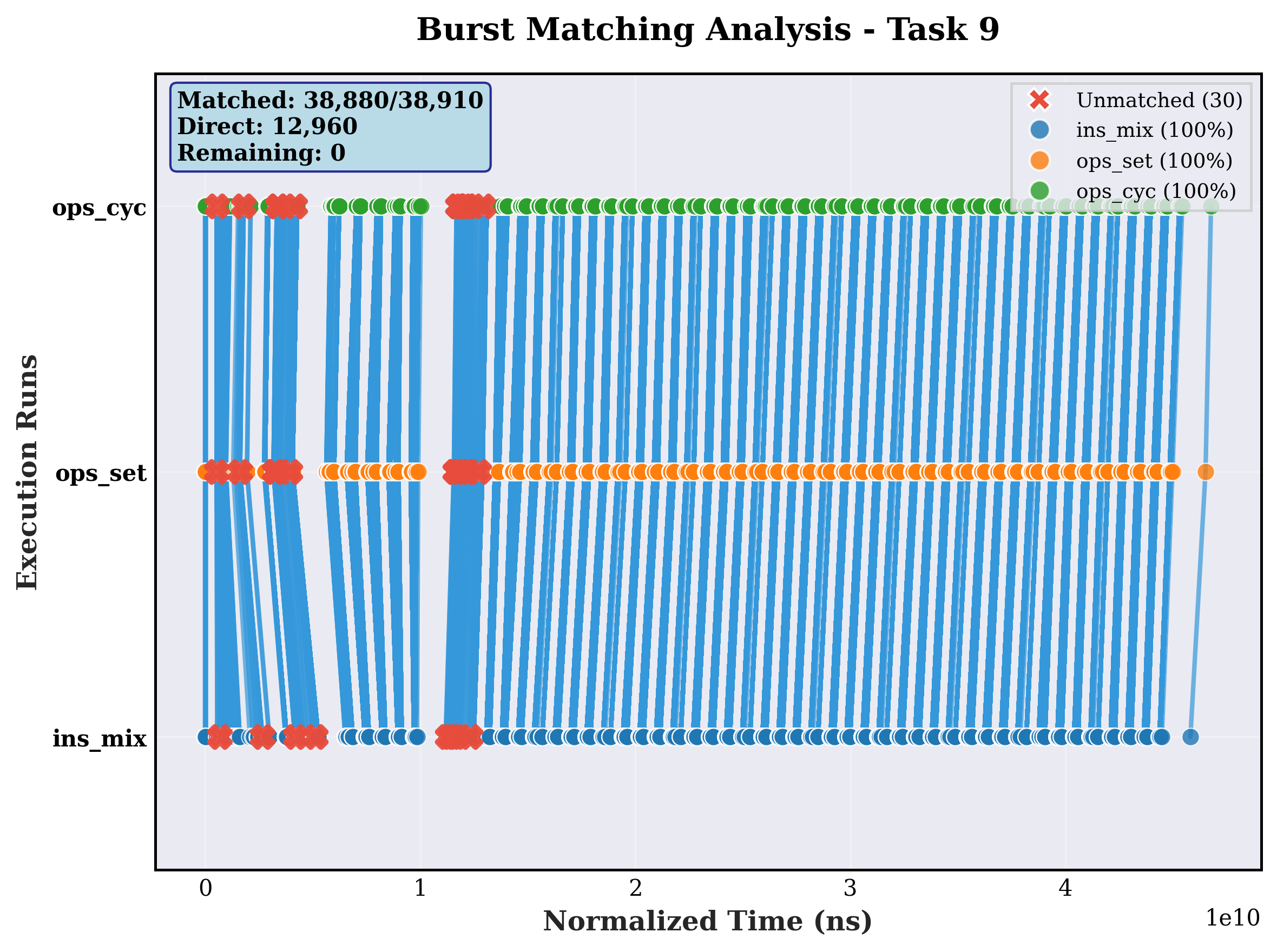}
\caption{\textbf{SOD2D} burst matching results after Stage 1 (pattern-based) with blue connecting lines showing direct matches. Red crosses show the 30 unmatched bursts requiring Stage 2 processing, concentrated in early temporal regions.}
\label{fig:SOD2D_stage1}
\end{figure}

Figure~\ref{fig:SOD2D_stage1} shows an example of \texttt{TaskId 9} results after Stage 1 (pattern-based matching), where 38,880 out of 38,910 bursts achieved correspondence, with 12,960 direct matches. The presence of red crosses indicates the 30 bursts that still require Stage 2 processing in this \texttt{TaskId}, a pattern consistently observed across almost all process executions.

\begin{figure}[h!tbp]
\centering
\includegraphics[width=1\columnwidth]{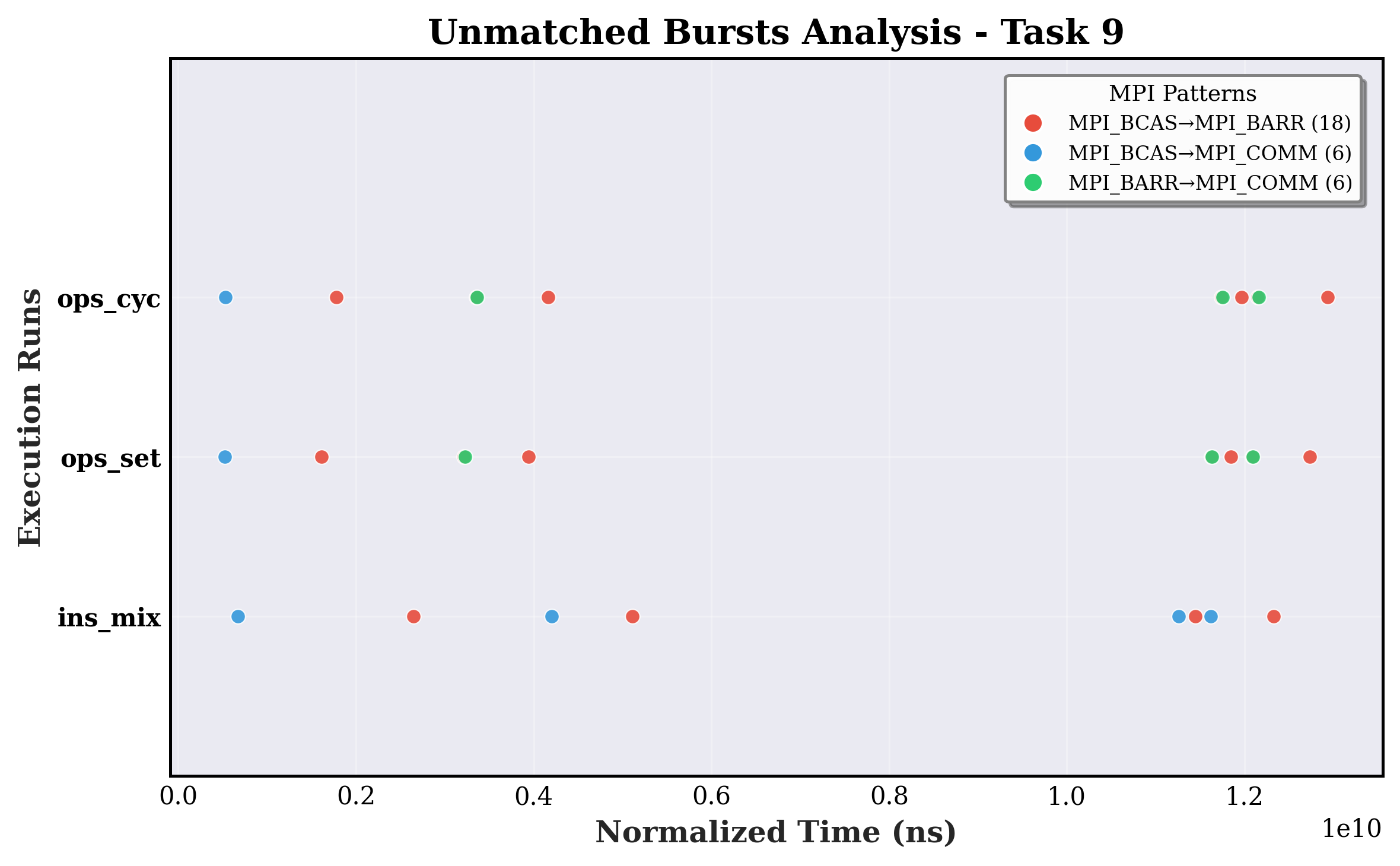}
\caption{Distribution of unmatched burst patterns in \textbf{SOD2D} after Stage 1. Three distinct \texttt{MPI\_BCAST} and \texttt{MPI\_BARRIER} communication patterns show timing-dependent behavior across execution runs.}
\label{fig:SOD2D_unmatched}
\end{figure}

Analysis of the unmatched patterns after Stage 1 (Figure~\ref{fig:SOD2D_unmatched}) reveals three distinct MPI communication patterns: \texttt{MPI\_BCAST} → \texttt{MPI\_BARRIER} (18 bursts), \texttt{MPI\_BCAST} → \texttt{MPI\_COMM\_FREE} (6 bursts), and \texttt{MPI\_BARRIER} → \texttt{MPI\_COMM\_FREE} (6 bursts). These 30 unmatched bursts are distributed across temporal regions and execution runs, primarily occurring during application initialization and collective communication phases.

\begin{figure}[h!tbp]
\centering
\includegraphics[width=1\columnwidth]{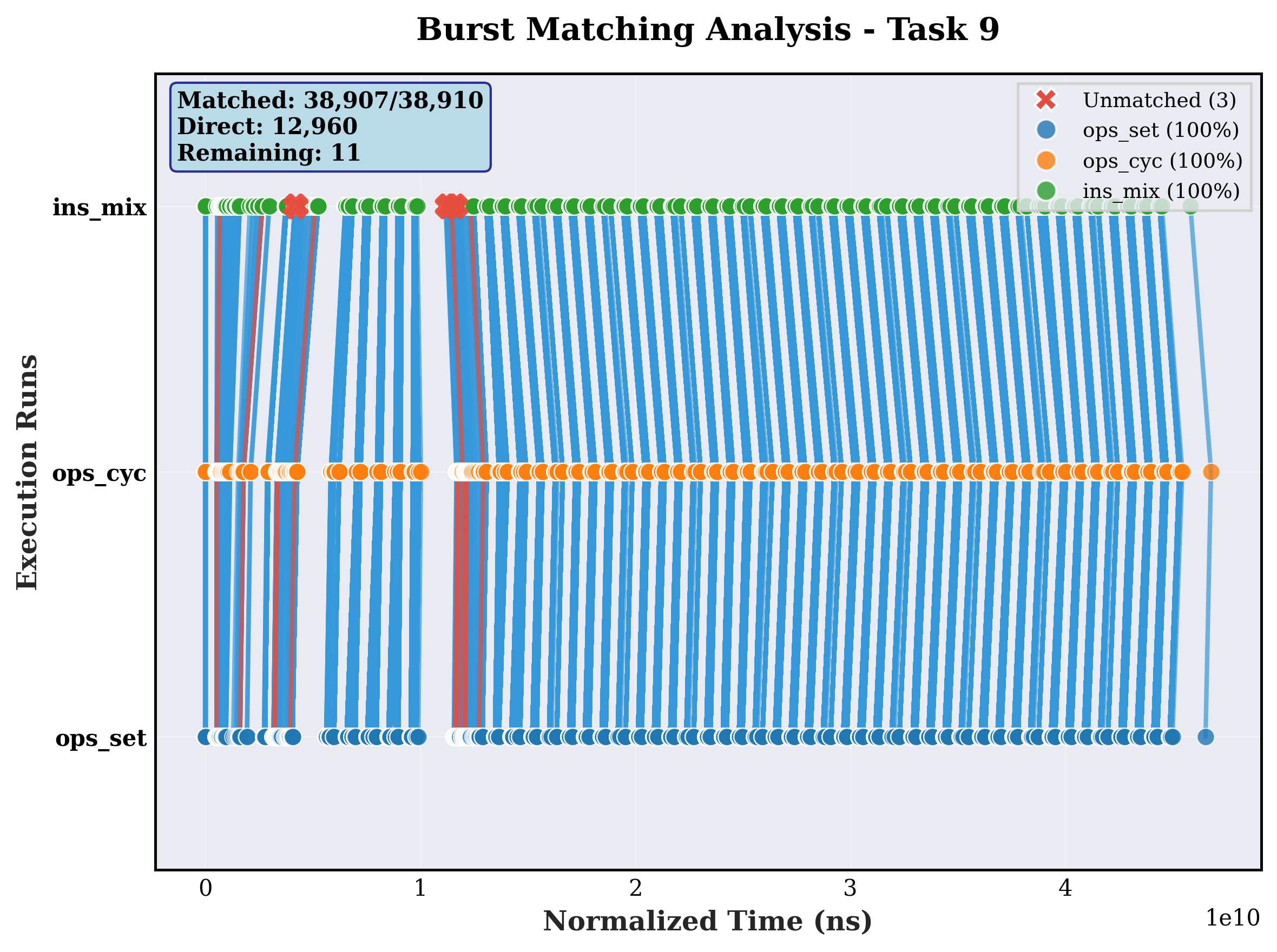}
\caption{\textbf{SOD2D} final matching results after Stage 2 structural constraints. Nearly perfect correspondence is achieved. Blue lines correspond to Stage 1 matching and red lines connecting bursts correspond to Stage 2 matching. Red crosses correspond to the three bursts belonging to \texttt{INS\_MIX} that remain unmatched.}
\label{fig:SOD2D_stage2}
\end{figure}

After applying Stage 2 structural constraints (Figure~\ref{fig:SOD2D_stage2}), the algorithm successfully matched 27 of the remaining 30 bursts, reducing unmatched bursts to only 3, achieving 99.99\% correspondence (38,880/38,910). The structural constraint algorithm effectively resolved the majority of remaining misalignments by respecting MPI collective communication boundaries and temporal locality within defined regions.

For the three counter sets, nearly complete matching was achieved: \texttt{INS\_MIX} reached 99.98\% (1,585,440/1,585,777), while both \texttt{OPS\_SET} and \texttt{OPS\_CYC} achieved 100.00\% matching rates.

\subsubsection{SeisSol: Complex Pattern Matching with Significant Non-Determinism}

\begin{figure}[h!tbp]
\centering
\includegraphics[width=1\columnwidth]{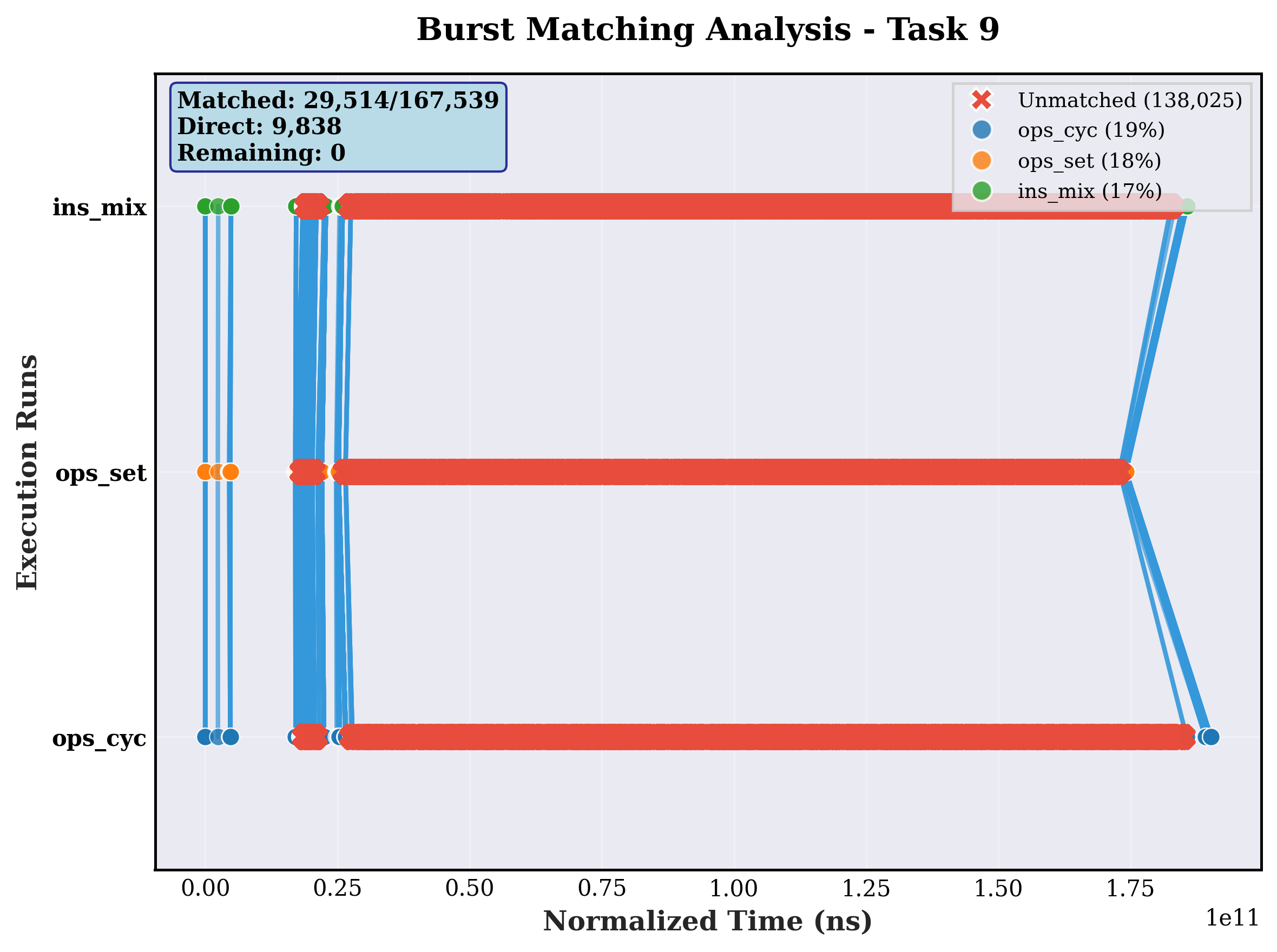}
\vspace{-0.4cm}
\caption{\textbf{SeisSol} burst matching results after Stage 1 showing significant execution non-determinism. Blue lines indicate successful correspondences in localized temporal regions, while red crosses condensed in the computational part represent 138,025 unmatched bursts requiring Stage 2 processing.}
\label{fig:seissol_stage1}
\end{figure}

SeisSol presented the most challenging matching scenario due to significant structural variations and the largest dataset volume ($>$ 5.5M bursts per execution). Figure~\ref{fig:seissol_stage1} shows \texttt{TaskId 9} results after Stage 1, where only 29,514 out of 167,539 bursts achieved correspondence (17.6\%), with 9,838 direct matches. The massive number of unmatched bursts (138,025) demonstrates complex non-deterministic behavior, with matching success varying across execution runs: \texttt{OPS\_CYC} (19\%), \texttt{OPS\_SET} (18\%), and \texttt{INS\_MIX} (17\%).

The major structural differences are concentrated in asynchronous communication patterns: \texttt{MPI\_IRECV} → \texttt{MPI\_IRECV} (up to 1,891 burst differences), \texttt{MPI\_ISEND} → \texttt{MPI\_ISEND} (1,751 differences), and \texttt{MPI\_TEST} → \texttt{MPI\_TEST} (1,746 differences). These reflect the application's heavy reliance on non-blocking communication with timing-dependent behavior.

Stage 2 structural constraints achieved substantial improvements, reducing \texttt{MPI\_ISEND} → \texttt{MPI\_ISEND} differences to 454 bursts and \texttt{MPI\_TEST} → \texttt{MPI\_IRECV} patterns to 134 bursts. The overall results show varying success: \texttt{INS\_MIX} achieved 99.99\% matching (5,492,274/5,492,883), while \texttt{OPS\_SET} and \texttt{OPS\_CYC} reached  90.11\% and 90.47\% due to larger burst counts and execution variations. The algorithm successfully handled 128 unique MPI patterns, demonstrating robustness under significant non-determinism environments.
\vspace{-0.3cm}

\vspace{-0.6cm}

\subsection{Matching Quality Validation}

To validate the accuracy of the burst matching algorithm, it was applied the methodology described in Section~\ref{subsec:evaluation_methodology} to multiple executions with identical counter sets, enabling controlled quantification of matching errors.

\begin{figure}[h!tbp]

\centering

\includegraphics[width=1\columnwidth]{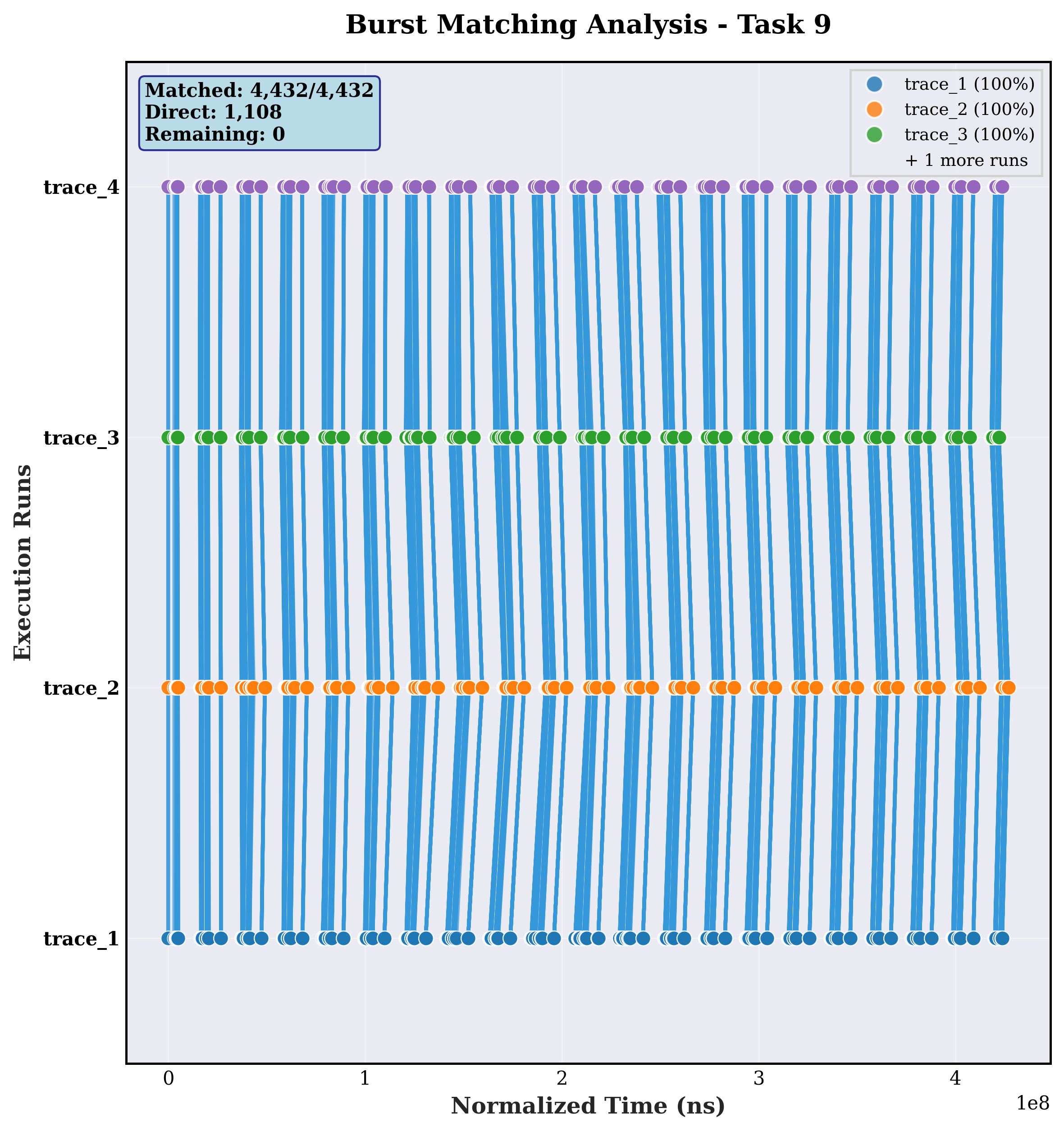}

\caption{Validation approach using \textbf{Lulesh} with identical counter sets. Blue lines connect corresponding bursts identified by the matching algorithm.}

\label{fig:lulesh_validation}
\end{figure}

For each application, results from 4 traces instrumented with the same counter set are compared\footnote{Except for SOD2D where only 2 traces per validation execution are available.}. For instance, Figure~\ref{fig:lulesh_validation} shows an example merge of 4 validation traces where for each matched burst, the first trace  is taken as baseline and relative differences are calculated from the other traces with respect to the first.




\subsubsection{Stream, Alya and Lulesh - Direct Matching Validation}

Analysis of validation for applications achieving perfect structural matching between executions, demonstrating deterministic behavior. Tables~\ref{tab:stream_instructions}, \ref{tab:alya_instructions}, and \ref{tab:lulesh_instructions} show  instruction counter precision across all applications, while cache hierarchy counters show  degradation from L1 to L3 levels, with L3 cache misses exhibiting in most of the cases the highest measurement variability despite perfect structural correspondence.

\begin{table}[h]
\centering
\caption{Stream \emph{INS\_MIX} Validation}
\label{tab:stream_instructions}
\begin{tabular}{|l|c|c|c|c|}
\hline
\textbf{Counter} & \textbf{Correlation} & \textbf{MAE} & \textbf{Rel Diff} & \textbf{$<$ 30\% Diff} \\
\hline
PAPI\_TOT\_INS & 1.000 & 60.922 & 0.019 & 99.9\% \\
PAPI\_SR\_INS & 1.000 & 9.460 & 0.011 & 100.0\% \\
PAPI\_LD\_INS & 1.000 & 23.629 & 0.017 & 100.0\% \\
PAPI\_BR\_INS & 1.000 & 5.785 & 0.005 & 100.0\% \\
PAPI\_L1\_DCM & 1.000 & 57.058 & 0.085 & 91.7\% \\
PAPI\_L2\_DCM & 0.997 & 122401.249 & 0.157 & 84.5\% \\
PAPI\_L3\_TCM & 0.926 & 12198.386 & 0.272 & 68.5\% \\
\hline
\end{tabular}
\end{table}


\begin{table}[h!]
\centering
\caption{Alya \emph{INS\_MIX} Validation}
\label{tab:alya_instructions}
\begin{tabular}{|l|c|c|c|c|}
\hline
\textbf{Counter} & \textbf{Correlation} & \textbf{MAE} & \textbf{Rel Diff} & \textbf{$<$ 30\% Diff} \\
\hline
PAPI\_TOT\_INS & 1.000 & 15.832 & 0.007 & 100.0\% \\
PAPI\_SR\_INS & 1.000 & 5.300 & 0.016 & 100.0\% \\
PAPI\_LD\_INS & 1.000 & 10.109 & 0.014 & 100.0\% \\
PAPI\_BR\_INS & 1.000 & 3.883 & 0.009 & 100.0\% \\
PAPI\_TOT\_CYC & 1.000 & 1214.215 & 0.045 & 99.3\% \\
PAPI\_L1\_DCM & 0.992 & 6.685 & 0.308 & 63.9\% \\
PAPI\_L2\_DCM & 1.000 & 3.164 & 0.207 & 74.1\% \\
PAPI\_L3\_TCM & 0.971 & 1.621 & 0.262 & 66.3\% \\
\hline
\end{tabular}
\end{table}

\vspace{-0.2cm}

\begin{table}[h]
\centering
\caption{Lulesh \emph{INS\_MIX} Validation}
\label{tab:lulesh_instructions}
\begin{tabular}{|l|c|c|c|c|}
\hline
\textbf{Counter} & \textbf{Correlation} & \textbf{MAE} & \textbf{Rel Diff} & \textbf{$<$ 30\% Diff} \\
\hline
PAPI\_TOT\_INS & 1.000 & 39.789 & 0.011 & 100.0\% \\
PAPI\_SR\_INS & 1.000 & 6.914 & 0.014 & 100.0\% \\
PAPI\_LD\_INS & 1.000 & 12.193 & 0.011 & 100.0\% \\
PAPI\_BR\_INS & 1.000 & 4.690 & 0.007 & 100.0\% \\
PAPI\_TOT\_CYC & 1.000 & 349.238 & 0.081 & 96.7\% \\
PAPI\_L1\_DCM & 1.000 & 12.651 & 0.204 & 76.3\% \\
PAPI\_L2\_DCM & 1.000 & 25.171 & 0.390 & 61.7\% \\
PAPI\_L3\_TCM & 0.973 & 11.387 & 0.499 & 51.6\% \\
\hline
\end{tabular}
\end{table}

\begin{figure}[h!tbp]
\centering
\includegraphics[width=1\columnwidth]{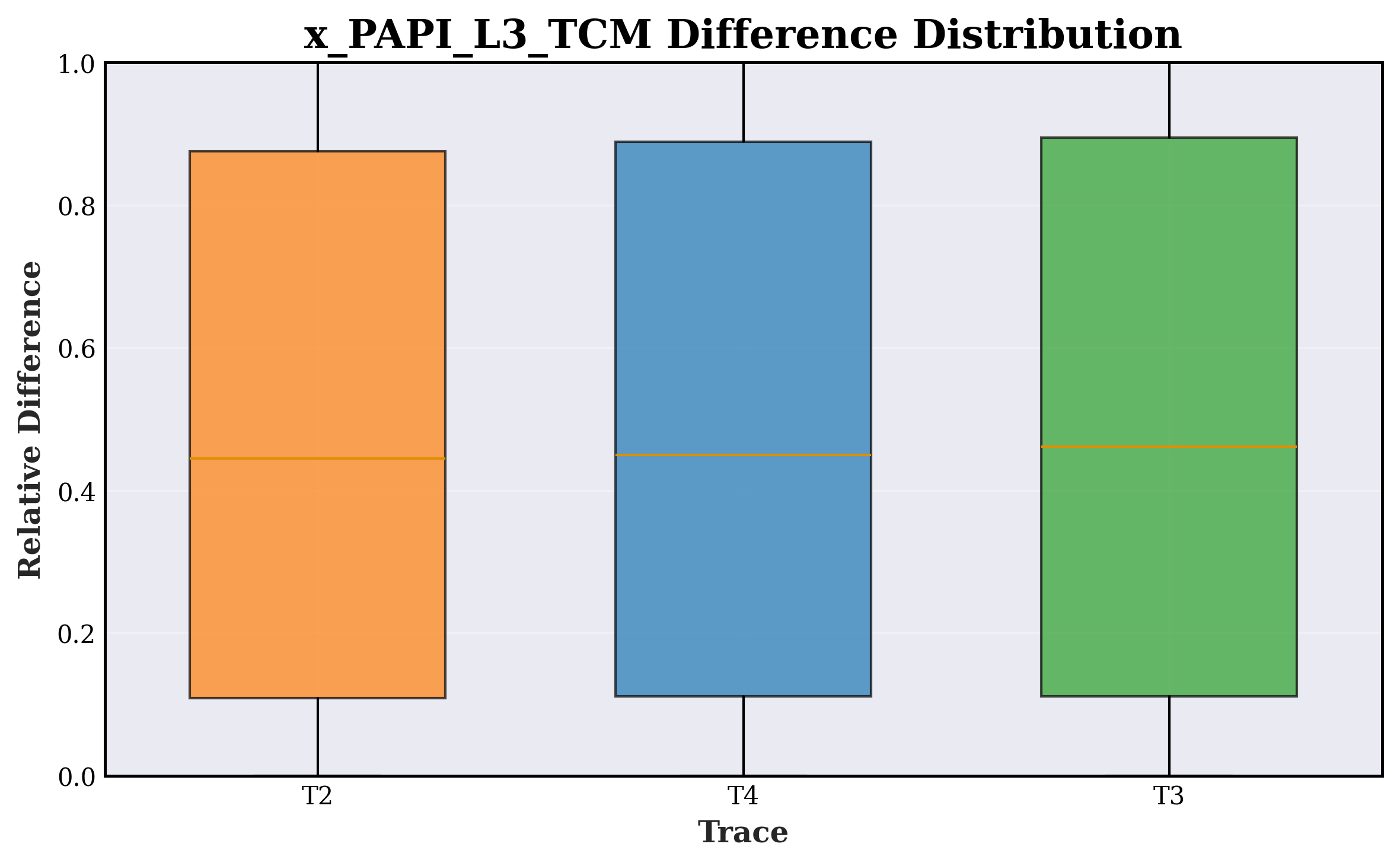}
\caption{L3 cache miss relative difference distribution for \textbf{Lulesh} validation. The boxplot shows relative differences of L3 cache misses across traces T2, T3, and T4 compared to baseline trace T1. The distributions show median values around 0.45-0.47 with interquartile ranges spanning approximately 0.10-0.87, indicating substantial variability in L3 cache miss measurements. All traces exhibit compressed distributions with upper whiskers extending to 1.0, reflecting measurement variations in L3 cache events despite identical computational phases.}
\label{fig:lulesh_l3_distribution}
\end{figure}

Figure~\ref{fig:lulesh_l3_distribution} illustrates the L3 cache validation challenge, showing substantial relative difference dispersion across traces despite strong correlations (0.973).

For the OPS\_SET and OPS\_CYC counter sets, detailed validation tables are not presented since these applications achieved perfect deterministic matching with no duplicate columns generated during trace fusion. All the operation counters demonstrated complete equivalence between executions; this deterministic behavior in floating-point counters reflects their consistent instruction scheduling across runs.  Only \texttt{PAPI\_TOT\_INS} and \texttt{PAPI\_TOT\_CYC} counters show the same minor variations documented in the INS\_MIX set analysis. 

\subsubsection{SOD2D - Pattern-Based Matching Validation}

SOD2D required pattern-based matching due to minor structural variations between executions, yet achieved 99.999\% matching success. Despite this near-perfect correspondence, the validation reveals distinct performance characteristics compared to direct matching applications. Table~\ref{tab:sod2d_validation} shows the validation results for SOD2D using the INS\_MIX counter set.

\begin{table}[h!]
\centering
\caption{SOD2D \emph{INS\_MIX} Validation Results}
\label{tab:sod2d_validation}
\begin{tabular}{|l|c|c|c|c|}
\hline
\textbf{Counter} & \textbf{Correlation} & \textbf{MAE} & \textbf{Rel Diff} & \textbf{$<$ 30\% Diff} \\
\hline
PAPI\_L2\_DCM & 1.000 & 2.729 & 0.512 & 50.8\% \\
PAPI\_L1\_DCM & 0.999 & 7.165 & 0.214 & 76.1\% \\
PAPI\_L3\_TCM & 0.999 & 2.000 & 0.488 & 51.3\% \\
PAPI\_SR\_INS & 0.963 & 7.387 & 0.010 & 99.9\% \\
PAPI\_LD\_INS & 0.876 & 13.693 & 0.009 & 99.7\% \\
PAPI\_BR\_INS & 0.875 & 6.258 & 0.005 & 99.7\% \\
PAPI\_TOT\_INS & 0.876 & 20.070 & 0.003 & 99.7\% \\
\hline
\end{tabular}
\end{table}

Unlike direct matching applications, SOD2D demonstrates inverted performance characteristics where cache-related counters (\texttt{L1\_DCM}, \texttt{L2\_DCM},\texttt{ L3\_TCM}) show excellent correlations (0.999-1.000) but significantly elevated relative differences (0.214-0.512), with only 50-76\% of measurements meeting the 30\% acceptance threshold. Conversely, instruction-based metrics display lower correlations (0.875-0.963) but excellent relative precision (0.003-0.010), with over 99\% of measurements within acceptable ranges. This pattern suggests that SOD2D's minor execution variations primarily manifest in cache behavior rather than computational instruction flow, likely due to memory access timing differences that don't affect core algorithmic execution but impact cache utilization patterns.

For the OPS\_SET and OPS\_CYC counter sets, SOD2D achieved perfect deterministic matching with no duplicate columns generated during trace fusion, similar to the direct matching applications. This demonstrates that floating-point operation counters maintain complete equivalence between executions even when structural matching is not direct. 

\subsubsection{SeisSol - Complex Non-Deterministic Validation}

SeisSol presents the most challenging validation scenario due to significant execution non-determinism. For the successfully matched bursts, the validation reveals varying precision across different counter types. Table~\ref{tab:seissol_validation} shows the validation results for SeisSol using the INS\_MIX counter set.

\begin{table}[h!]
\centering
\caption{SeisSol \emph{INS\_MIX} Validation}
\label{tab:seissol_validation}\begin{tabular}{|l|c|c|c|c|}
\hline
\textbf{Counter} & \textbf{Correlation} & \textbf{MAE} & \textbf{Rel Diff} & \textbf{$<$ 30\% Diff} \\
\hline
PAPI\_SR\_INS & 0.994 & 4.942 & 0.015 & 97.2\% \\
PAPI\_LD\_INS & 0.993 & 11.373 & 0.015 & 97.3\% \\
PAPI\_TOT\_INS & 0.993 & 41.992 & 0.019 & 97.3\% \\
PAPI\_L1\_DCM & 0.991 & 4.738 & 0.669 & 38.5\% \\
PAPI\_BR\_INS & 0.989 & 4.725 & 0.008 & 97.4\% \\
PAPI\_L2\_DCM & 0.980 & 0.885 & 0.092 & 87.2\% \\
PAPI\_L3\_TCM & 0.976 & 1.129 & 0.174 & 75.7\% \\
\hline
\end{tabular}
\end{table}

Despite the application's complexity, SeisSol demonstrates robust validation for successfully matched bursts. SeisSol exhibits strong correlations across all counter types (0.976-0.994), with instruction-based counters maintaining excellent precision, over 97\% of measurements fall within the 30\% acceptance threshold with relative differences below 0.019. Cache counters show progressively degraded precision from L3 (75.7\% acceptance) to L1 (38.5\% acceptance), with L1 cache misses displaying the highest relative differences (0.669). This validation pattern indicates that while SeisSol's execution exhibits significant structural variations requiring sophisticated matching algorithms, the core computational phases remain sufficiently similar to enable meaningful performance analysis, with instruction flow showing greater consistency than memory hierarchy behavior.

\begin{figure}[h!tbp]

\centering

\includegraphics[width=1\columnwidth]{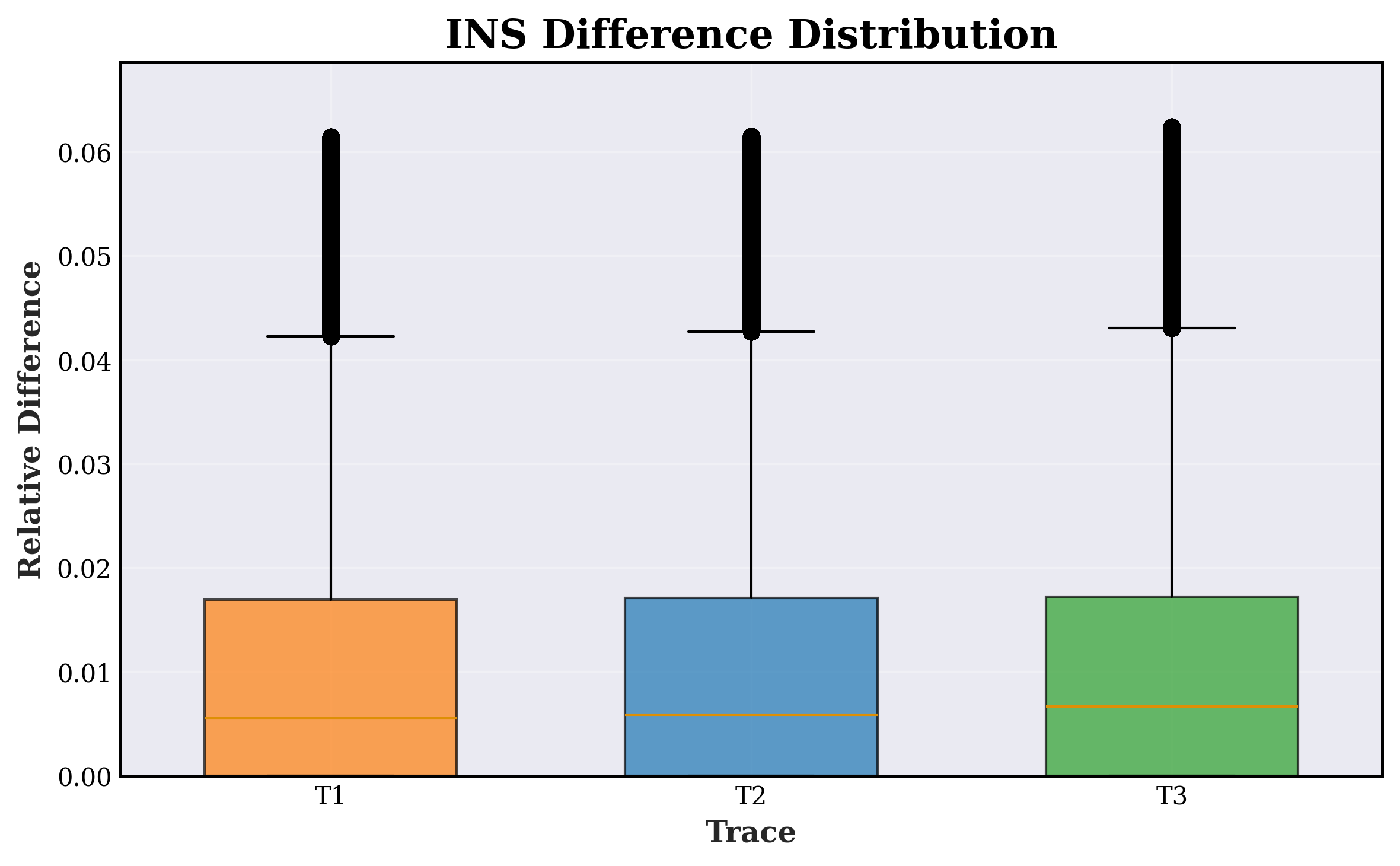}

\caption{Boxplot showing the relative difference distribution of the total instructions hardware counter  for \textbf{SeisSol} validation.}

\label{fig:seissol_ins_distribution}
\end{figure}

Figure~\ref{fig:seissol_ins_distribution} illustrates the instruction counter validation challenge, showing relative difference distributions across traces T1, T2, and T3 compared to the baseline. All three traces display remarkably consistent patterns with medians around 0.005-0.006 and tight interquartile ranges (0.0-0.017), indicating excellent matching quality for instruction counts. The distributions show minimal outliers extending only to approximately 0.062, demonstrating that instruction counter measurements achieve highly reliable correspondence across traces with very low variability.

For the OPS\_SET  and OPS\_CYC counter validation, SeisSol demonstrates remarkable stability in floating-point operations despite its complex execution patterns. Table~\ref{tab:seissol_ops_validation} shows these results, where counters are no longer equivalent as in previous applications due to the non-deterministic nature requiring duplicate column generation.


\begin{table}[h!]
\centering
\caption{SeisSol \emph{OPS\_SET} Validation Results}
\label{tab:seissol_ops_validation}
\begin{tabular}{|l|c|c|c|c|}
\hline
\textbf{Counter} & \textbf{Correlation} & \textbf{MAE} & \textbf{Rel Diff} & \textbf{$<$ 30\% Diff} \\
\hline
PAPI\_VEC\_SP & 1.000 & 0.000 & 0.000 & 100.0\% \\
PAPI\_SP\_OPS & 1.000 & 0.000 & 0.000 & 100.0\% \\
PAPI\_VEC\_DP & 0.993 & 0.000 & 0.000 & 99.8\% \\
PAPI\_VEC\_INS & 0.993 & 0.000 & 0.000 & 99.8\% \\
PAPI\_DP\_OPS & 0.993 & 0.000 & 0.000 & 99.4\% \\
PAPI\_FP\_OPS & 0.993 & 0.000 & 0.000 & 99.4\% \\
PAPI\_FP\_INS & 0.992 & 0.000 & 0.000 & 99.4\% \\
\hline
\end{tabular}
\end{table}

Single precision operations (\texttt{PAPI\_VEC\_SP},\texttt{ PAPI\_SP\_OPS}) achieve perfect correlations (1.000) with zero relative differences and complete acceptance rates (100.0\%). Double precision and vectorization counters maintain excellent correlations (0.992-0.995) with near-zero relative differences (0.000) and acceptance rates between 99.4-99.8\%. This  matching precision in floating-point counters confirms that these metrics can be confidently merged across different executions without introducing measurement errors, as they exhibit virtually no variability between matched executions despite SeisSol's non-deterministic behavior.

\section{Conclusion and Future Work}

\label{sec:conclusion}

This work presents a heuristic-based methodology for merging HPC execution traces to extend hardware counter coverage. Our analysis across five representative applications reveals distinct behavioral patterns that determine matching complexity and success rates.

The evaluation identifies two primary application classes. Deterministic applications (Stream, Alya, Lulesh) exhibit consistent MPI communication structures across executions, enabling direct matching with 100\% correspondence rates. Non-deterministic applications (SOD2D, SeisSol) require advanced pattern-based matching and structural constraint algorithms to achieve near-perfect matching rates in at least one of all executions merged.

Validation results establish clear reliability rankings across hardware counter categories. Instruction and floating-point operation counters demonstrate  consistency with perfect correlations (1.000) across deterministic applications, maintaining stability even in complex applications like SeisSol. This demonstrates that floating-point counters can be confidently merged from separate executions without introducing measurement errors, providing substantially expanded feature spaces.

However, cache hierarchy counters exhibit systematic measurement challenges with substantial variability patterns. L1 cache misses show the highest relative differences, while L3 cache misses demonstrate moderate correlations (0.926-0.999) but elevated relative differences (0.174-0.499) across applications. Acceptance rates for cache counters range from 38.5\% (L1) to 87.2\% (L2), and the fact that this is observed also in direct-matching scenarios, indicates inherent timing-dependent measurement sensitivity rather than algorithmic limitations at the burst matching phase. This findings suggest that cache misses counters may not be a good fit for a dataset built for compute bursts performance prediction.

The methodology successfully enables hardware counter coverage expansion, with over 97\% of successfully matched bursts achieving acceptable correspondence for instruction-based metrics. Even in challenging non-deterministic scenarios, floating-point counters maintain perfect precision, providing substantial feature space enrichment for performance analysis applications.
\vspace{-0.4cm}

\subsection{Future Work}
\vspace{-0.2cm}
\textbf{Cache Counter Investigation}: The variability in cache hierarchy measurements requires detailed analysis using Paraver. Future work should identify and study specific computational phases generating large cache miss differences.

\textbf{ML Model Expansion}: The validated synthetic traces can enable training performance prediction models on expanded feature spaces and expand our prior work on performance analysis of HPC workloads. Comparative studies should evaluate prediction accuracy improvements between limited counter sets and the enriched datasets, quantifying the impact of extended hardware counter coverage on model generalization and performance. Future work on this direction should also integrate the learning on counter variability across runs in order to build better ML models. 

\textbf{Production Deployment}: Production deployment requires evaluation across broader application domains and HPC architectures to confirm generalization and identify architecture-specific adaptations needed for optimal matching performance.

\section*{Appendix}
\label{sec:appendix}
\begin{algorithm}[H]
\caption{Primary Burst Matching (Direct + Pattern-Based)}
\label{app:alg}
\begin{algorithmic}[1]
\REQUIRE A set of executions $\mathcal{E} = \{E_1, E_2, \dots, E_n\}$, where each $E_i$ is a trace with $R$ MPI ranks: $E_i = \{e_i^1, e_i^2, \dots, e_i^R\}$

\FOR{each MPI rank $r \in \{1, \dots, R\}$ (in parallel)}
    \STATE Let $k_i^r$ be the number of computation bursts for rank $r$ in execution $E_i$
    \STATE Extract the burst sequences $B_i^r = \{b_{i,1}^r, \dots, b_{i,k_i}^r\}$ from each  $E_i$

    \IF{$k_1 = k_2 = \dots = k_n$ and MPI sequences are identical $\forall i$}
        \FOR{$j = 1$ to $k_1$}
            \STATE Assign \texttt{burst\_id} $= \text{``}r\_\text{d}\_j\text{''}$ to $b_{i,j}^r$ for all $i$
        \ENDFOR
    \ELSE
        \STATE Compute the set of patterns $P = \{p = (\texttt{MPI}_{\text{before}}, \texttt{MPI}_{\text{after}})\}$
        \STATE Filter $P_{\text{det}} \subseteq P$ such that $p \in P_{\text{det}} \iff \text{freq}_{E_1}(p) = \dots = \text{freq}_{E_n}(p)$
        \FOR{each $p \in P_{\text{det}}$}
            \STATE Let $f = \text{freq}_{E_i}(p)$ (same for all $i$)
            \FOR{$j = 1$ to $f$}
                \STATE Assign \texttt{burst\_id} $= \text{``}r\_\text{p}\_p\_j\text{''}$ to $j$-th occurrence of $p$ in all $E_i$
            \ENDFOR
        \ENDFOR
    \ENDIF
\ENDFOR
\end{algorithmic}
\end{algorithm}

%
%
%

\bibliographystyle{spmpsci}

\bibliography{references}

\end{document}